\documentstyle[11pt,epsfig]{article}

\def\laq{~\raise 0.4ex\hbox{$<$}\kern -0.8em\lower 0.62 
ex\hbox{$\sim$}~}
\def\gaq{~\raise 0.4ex\hbox{$>$}\kern -0.7em\lower 0.62 
ex\hbox{$\sim$}~}

\begin{document}
\begin{titlepage}

\begin{flushright}
CERN-TH/2003-304 \\
hep-th/0312182
\end{flushright}

\vspace*{1cm}

\begin{center}
\Large{\bf  A Model for the Big Bounce}

\vspace*{1cm}

\large{  G. Veneziano}
\bigskip
\normalsize

{\sl  Theoretical Physics Division, CERN,
CH-1211 Geneva 23, Switzerland}\\

\vspace*{5mm}

\begin{abstract}
 I motivate a proposal for modeling, at weak string coupling,  the ``Big Bounce" transition from a growing-curvature phase to standard (FRW) cosmology in terms of a pressure-less dense gas of  ``string-holes" (SH),  string states lying on the  correspondence curve between strings and black holes.  During this phase SH evolve in such a way that  temperature and (string-frame) curvature remain  $O(M_s)$ and  (a cosmological version of) the holographic entropy bound remains saturated. This reasoning also appears to imply a new interpretation of the Hagedorn phase transition in string theory.

\end{abstract}
\end{center}
\end{titlepage}

The fate of the big bang singularity of General Relativity (GR)  is one of the most important unsolved  problems in string theory.
Unlike time-like singularities,  space-like curvature singularities --such as those related to the big bang or to the interior of black holes-- confront string theory with  high-curvature, strongly time-dependent backgrounds, a {\it terra incognita}, to a large extent. 

Understanding black hole thermodynamics in string theory from a statistical mechanics viewpoint has been comparatively  easier (for a review see e.g. \cite{BHentropy} ). For extremal and near-extremal black holes analytic treatments are possible and reproduce amazingly well GR expectations. For more usual (e.g. Schwarzschild-like) black holes an analytic treatment is not available but some understanding has been gathered on how string states progressively turn into black holes as the string coupling is increased. In particular, along a correspondence curve (see Fig. 1) given by  \cite{BHScorr}:
\begin{equation}
M = M_s  g_s^{-2} \; ,
\label{correspline}
\end{equation}
where , $M_s = l_s^{-1}$ and $g_s$ are the string mass and string coupling, respectively, properties of strings and  black holes match impressively well \cite{matching}. This correspondence also supports an earlier conjecture \cite{Bowick} about the late stage of the evaporation process of black holes in string theory. Such a process inevitably takes a black hole to the correspondence curve (\ref{correspline}) where it turns into a string and typically decays into (almost) massless quanta, thus avoiding any singularity at the end of its life.
\begin{figure}
\centerline{\epsfxsize = 10cm  \epsffile{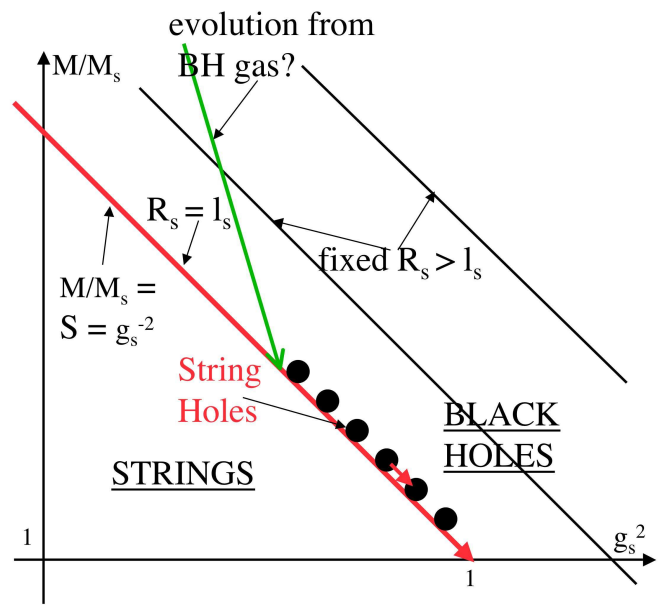}}
\vskip 2mm
\caption[a]{Black-hole/string phase diagram in a log--log $g_s$ vs. $M/M_s$ plot.}
\label{FIGURE1} 
\end{figure}
In this note I will argue that the above understanding of black holes  can be used to make progress on the issue of the cosmological singularity  or, better, on how to describe the transition from a growing curvature phase to a FRW cosmology, transition that I would like to dub  the ``Big Bounce". The link between these two seemingly unrelated issues is based on  the use of of holography and of related cosmological entropy bounds .

Following pioneering work by Bekenstein \cite{Bek}, several cosmological entropy bounds have been proposed and (theoretically) tested during the past few years. At the level at which we shall be working here, they can be simply stated \cite{FS}, \cite{HB} as bounding the entropy of a large region of space by that of a closely packed gas of black holes whose (Schwarzschild) radius coincides with the Hubble radius $H^{-1}$. Neglecting throughout numerical factors, this leads to an entropy density bound $\sigma_{max}$ given by \footnote {We shall work troughout in a generic number $D$ of space-time dimensions and set $\hbar = c =1$.}:
\begin{equation}
\sigma_{max} = \frac{S_{max}}{V} =  H  M_P^{D-2}
\label{Hbound}
\end{equation}
 $M_P$ denotes the $D$-dimensional Planck mass, the inverse of the $D$-dimensional Planck length, $l_P$.

In a  Universe dominated by a perfect fluid of equation of state $p = w \rho$ this bound can be compared to the actual entropy of the fluid:
\begin{equation}
\sigma_F = \frac{p + \rho}{T} = (1+w)  \frac{H^2 ~ M_P^{D-2}}{T}  =  \frac{(1+w) H}{T} ~ \sigma_{max}
\label{fluid}
\end{equation}
In standard (post-big bang) cosmology one has typically $H/T \ll 1$ and, furthermore, the bound is better and better fulfilled as the Universe expands since $H/T$ decreases with time for $w <1$ (and remains constant for $w=1$).
In pre-big bang cosmology \cite{PBB}, as well as in other scenarios \cite{EKP} where a growing-curvature phase precedes the big bang,
the situation is completely reversed. A stiff fluid, $w=1$, is the only safe case, while for $w <1$ the bound is increasingly threatened  as the big bang is approached. Clearly we have to understand what happens as we approach the limit $T \rightarrow H$.

In this limit another phenomenon becomes relevant: particle production from the time-dependent geometry or, if we prefer, the amplification of quantum fluctuations in a time-dependent background geometry. This process is far from being adiabatic and produces an entropy density 
 given by \cite{SQ}:
\begin{equation}
\sigma_Q = g_* H^{D-1} =   g_*~  \left( \frac{H}{M_P} \right)^{D-2} ~\sigma_{max} 
\label{quantum}
\end{equation}
where $g_*$ is the effective number of species that are produced. Note that the above entropy density is, parametrically speaking, that of a relativistic gas at  $T \sim H$ with $g_*$ species.

Let us now consider the  three entropy densities (\ref{Hbound}), (\ref{fluid}),  (\ref{quantum}) in a string theory context assuming $T$ to be bounded by
the Hagedorn temperature, $T \le T_H \sim M_s$ and curvature to be bounded by $l_s^{-2}$. Let us also consider (very) weakly coupled string theory, $g_s^2 = {\rm e}^{\phi} \ll 1$ and recall the (weak-coupling, closed string) relation:
\begin{equation}
\left( \frac{M_s}{M_P} \right)^{D-2} = g_s^2 = {\rm e}^{\phi} \; ,
\label{MSMP}
\end{equation}
where $\phi$ is the dilaton.
In a string cosmology context the dilaton evolves in time and so does, through (\ref{MSMP}), the ratio of the Planck to the string mass. Furthermore, in the pre-big bang scenario the string coupling starts extremely small and grows towards the bounce while the opposite is true in the ekpyrotic/cyclic Universe. Hereafter we shall concentrate on the former case.

Can we saturate the entropy bound for $T \ll M_s$?. The answer is  apparently negative. This looks quite obvious for (\ref{quantum}) since $(H/M_P)^{D-2} < (M_s/M_P)^{D-2} = g_s^2 \ll 1$ If we keep $g_*$ fixed as we decrease the coupling there is no way to saturate the bound from $\sigma_Q$. We can argue similarly for $\sigma_F$ since $H/T$ must go to zero as we turn off Newton's constant (i.e. the string coupling).
This naive failure to saturate the bound, however, suggests by itself a way to achieve saturation even at very small string coupling: we need the effective number of species $g_*$ to grow as the string coupling is decreased. But how? Actually, we do know already of states that have  the desired property: they are precisely the (marginal) black holes that lie on the above-mentioned correspondence curve of Eq. (\ref{correspline}) (and Fig. 1) that we now complete as: 
\begin{equation}
M = M_s  g_s^{-2} = M_s ~ {\rm e}^{-\phi} = M_P ~ \exp{\left(-\frac{D-3}{D-2}\phi\right)}
\label{corresp.curve}
\end{equation}
The Schwarschild radius and Hawking temperature of these black holes are given by the string scale and mass, respectively: we will refer to them in the following as string-holes (SH).
The entropy of a SH is  given equivalently by the Bekenstein-Hawking or by the weak-coupling string formulae \cite{BHScorr}:
\begin{equation}
S_{SH} = (l_s /l_P)^{D-2} =  g_s^{-2} = S_{string} \equiv M l_s \; ,
\label{SH entropy}
\end{equation}
The possibility of forming small primordial black holes in the pre-bang phase of string cosmology, as a result of
the blue spectra of tensor and scalar perturbations characteristic of these scenarios \cite{PBB}, was
already  emphasized
in \cite{CLLW}, where a possible connection between primordial black hole formation and observation of a stochastic gravitational-wave background at (advanced versions of present)  interferometers was also pointed out.

Let us check now that the identification of the final stage of collapse with such a gas of SH satisfies other consistency checks and let us also argue for an evolution of such a gas once it is formed.
The following quantities (referring to the string frame) can be readily computed:
\begin{eqnarray}
\label{SHgas}
\rho_{SH} &=&  M_s~g_s^{-2}/ l_s^{D-1} = {\rm e}^{- \phi} ~ M_s^D \, ,\nonumber \\
\sigma_{SH} &=&  \left(  l_s/l_P \right)^{D-2}/   l_s^{D-1} =  {\rm e}^{- \phi} ~ M_s^{D-1}    \; ,\nonumber  \\
\end{eqnarray}
and since $H \sim T  \sim M_s$,    we immediately check that $\sigma = \rho/T$ and that the entropy bound is saturated.

In order to derive an evolution equation for the gas we impose the 2nd law and demand that  entropy in a fixed comoving volume (always in the string frame) does not decrease. Since
\begin{equation}
S = \sigma_{SH} V \sim  {\rm e}^{- \phi} \, V ~ \sim  {\rm e}^{-\bar{ \phi}}  \; ,
\label{Scons}
\end{equation}
we  see that, necessarily, $\dot{\bar{\phi}} \le 0$ with the equality sign holding true for an adiabatic evolution.
In (\ref{Scons})  we have introduced the standard ``shifted dilaton" field $ \bar{\phi}  = \phi - 1/2 \log( g) + {\rm const.}$ that plays an important role in string cosmology since it is invariant under scale-factor duality transformations and their continuous generalizations.
This result was already obtained some years ago \cite{GSL} in a less specific framework and is very encouraging since it means that the strong-curvature evolution takes us towards a FRW Universe.

In the present case we expect the evolution of the gas to be almost adiabatic, al least for a while, in which case $\dot{\bar{\phi}} = 0$ and the picture becomes particularly appealing: each SH loses mass while the coupling grows in order to stay on the correspondence curve. The mass that is lost  turns out to be {\it precisely} what is needed to form more SH's so that they keep  filling the growing number of Hubble volumes while remaining closely packed. In the string frame the equation of state is  that of dust  ($p=0$) since $\rho \sim \exp(-\phi) \sim V^{-1}$. This looks natural since SH are non-relativistic at small $g^2$. Furthermore,   $\dot{\bar{\phi}} = 0$ and $p=0$ are both invariant under T-duality transformations, suggesting that, indeed, the SH-phase can serve as bridge between two duality-related pre and post-bounce cosmologies.

We can also ask whether such a phase can be the solution of the equations of motion of a string theoretic effective action.
We  claim that this is indeed the case. Following \cite{GMV}, consider the tree-level effective action of (say heterotic) string theory including all-order $\alpha'$ corrections (recall that we are interested in a high-curvature small-coupling regime)  but add to it a ``matter" piece  providing the contribution of the dense gas of SH's to the stress-energy tensor.
Generalizing slightly the arguments given in \cite{GMV} we can easily show that a solution with $\dot{\bar{\phi}} = 0$, $p=0$, and $\rho \sim \exp{(-\phi)} \sim V^{-1}$  exists whenever a certain $\beta$-function, $\beta(H^2)$ with $\beta(0)=0$, has (at least) one non-trivial zero. This is the case, for instance, in the explicit model discussed in \cite{GMV}. The location of the smallest non-trivial zero and the derivative of $\beta$ at that point determine the Hubble parameter and $~ \rho\exp{(\phi)} $ during the SH phase, respectively. The latter quantity turns out to be automatically of the right order to correspond to our dense SH gas.

A SH condensate would then be the physical description of the so-called ``string phase" of (weak-coupling) PBB cosmology, the horizontal evolution in Fig. 2, during which the string-frame Hubble parameter is fixed (giving de Sitter like evolution) while the dilaton keeps growing at a rate determined by the Hubble parameter via $\dot{\phi} = (D-1) H$.
In the absence of matter sources such a solution does not exist, generically, but, as argued above, the opposite is true if we allow for  SH production as the Hubble parameter approaches the string scale.
In the case of ``strongly-coupled" PBB cosmology, instead, the entropy bound is already saturated at weak curvature when the dilaton reaches a critical value (see again Fig. 2) owing to production of light quanta including $D-0$ branes. This complementary situation was described several years ago by Maggiore and Riotto \cite{MR}.
\begin{figure}
\centerline{\epsfxsize = 10cm  \epsffile{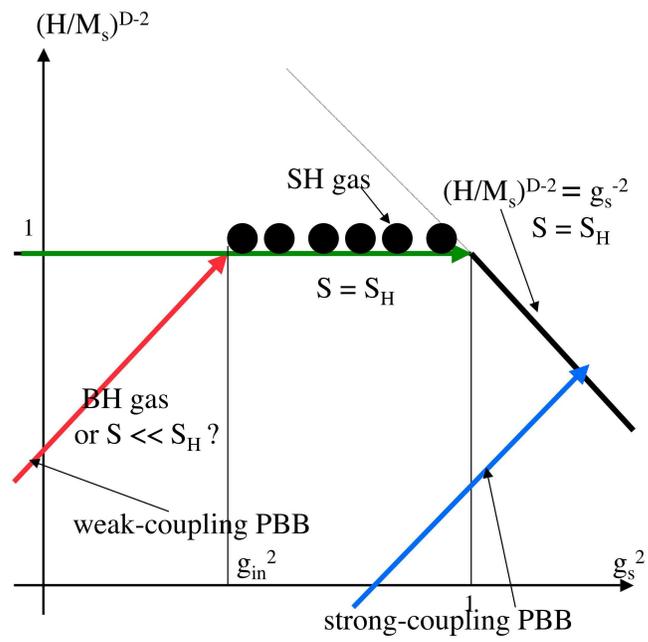}}
\vskip 3mm
\caption[a]{Pre-big bang phase diagram in a log--log $g_s$ vs. $H/M_s$ plot.}
\label{FIGURE2} 
\end{figure}
Let us now compare our model with the ``Black Crunch" picture recently considered in \cite{BF1}. Also there one wishes to keep the bound on entropy saturated and time-independent (for a fixed comoving region of space). This imposes:
\begin{equation}
S_{max} = V H M_P^{D-2} = {\rm  const.}
\end{equation}
In \cite{BF1} it was argued that saturation can be achieved even at sub-string curvature scales.
This implies  $V~M_P^{D-2} \sim H^{-1} \sim t $ and selects $p=\rho$ as the relevant equation of state in the Einstein frame. It is straightforward to convert our solution from the string to the Einstein frame using standard relations.
One finds that  the Einstein-frame metric and Hubble parameter are related to each other and to the E-frame cosmic time by 
\begin{equation}
H_E \sim 1/\sqrt{g_E}  \sim t_E^{-1} \; ,
\end{equation}
hinting again at a stiff equation of state. Using the appropriate E-frame conservation equation (see, e.g. Eq. (2.117) of \cite{PBB}),  and after inserting our solution for $\phi$, one finds indeed the conservation equation for a $p=\rho$ fluid. 

Although we have argued  that it is unlikely for the entropy bound to be saturated {\it before} the string scale is reached this is far from being obvious under all circumstances (e.g. for particularly entropic initial states or as a consequence of some highly  non-adiabatic classical process). What happens then if the bound becomes saturated at $H \ll M_s$?
Since the Hubble radius is shrinking, it is obvious that the evolution of the system will follow trajectories in the $M-g_s^2$ plane that cross the constant $g_s^2 M$ (i.e. constant Schwarzschild radius)  lines {\it downward} (see Figure 1). 
Thus, eventually, the evolution will take us to the correspondence line and will follow it thereafter  (as long as the string coupling is weak). The effective E-frame equation of state will not change.
We conclude that the occurrence of the SH phase is a generic outcome of pre-bang evolution in string theory. Indeed, it is likely to be also the final outcome of the chaotic BKL evolution that was argued \cite{BKL} to occur generically in low-curvature small-coupling string or M-theory cosmology.

What is the ultimate fate of the SH phase we have advocated? The answer is not obvious: as soon as the string coupling reaches values $O(1)$, $g_*$ too  becomes  $O(1)$ and there is no longer a way to keep the bound saturated while the curvature stays constant. If the dilaton keeps growing one could imagine connecting smoothly to the regime described in \cite{MR} staying on the rightmost boundary line in Fig. 2. If instead the dilaton (or rather the ratio $M_P/M_s$) freezes various possibilities emerge: the SH gas could disintegrate immediately into radiation; it could keep expanding adiabatically, following the $w=1$ evolution advocated in \cite{BF2}; or it could expand with considerable entropy release and $w <1$, since in  this case the bound itself grows in time.  In any case, ultimately, decay into radiation should take place and the equation of state  will become $p = \rho/(D-1)$.
  This does not mean, however, that there is no memory of what the actual phase preceding the bounce was. Properties of that phase would  leave an imprint on large-scale perturbations that should still survive in the SH gas and in the radiation that follows it. Hopefully, these perturbations will have the right properties to produce the observed large-scale structure of our Universe.

Quite apart from cosmology, our  considerations also appear to solve a more general puzzle within string theory. Supposedly, both temperature and (string frame) curvature  are bounded by one and the same scale, $M_s$. However, temperature is related to energy density $\rho$ while going from $\rho$ to curvature involves Newton's constant which can be made arbitrarily small by going to the weak-coupling limit. Hence the two bounds appear to be parametrically different. Our point is that they  can be simultaneously saturated if the number of effective degrees of freedom near the Hagedorn temperature grows as $G_N^{-1} \sim g_s^{-2}$ i.e. precisely like the entropy of SH's.

This  also suggests a novel interpretation of the limiting temperature of string theory. As one increases $T$ at small non zero string coupling,  competition between Boltzmann suppression and entropy should lead, already at  $T = T_H\left( 1- O(g_s^2)\right)$,   to a phase transition  whereby a dense gas of SH's of mass $M_s/g^2$ replaces the more conventional low-temperature gas of light strings/particles. At this point, however, gravitational back-reaction will become important and will cause the proper volume of the gas to become time-dependent. Thus only a cosmological interpretation of the Hagedorn phase transition will be possible.

This work was carried out mostly during  the Superstring Cosmology Workshop held at the Kavli Institute for Theoretical Physics, University of California at Santa Barbara (August -- December 2003). I wish to thank many of the participants in the workshop, and in particular
  T.Banks, R. Brandenberger,  R. Brustein and M. Gasperini  for useful discussions and encouragement.

\end{document}